\documentclass{article}
\usepackage{arxiv}
\usepackage[utf8]{inputenc} 
\usepackage[T1]{fontenc}    
\usepackage{hyperref}       
\usepackage{url}            
\usepackage{booktabs}       
\usepackage{amsfonts}       
\usepackage{nicefrac}       
\usepackage{microtype}      
\usepackage{lipsum}		
\usepackage{graphicx}
\usepackage{mathrsfs}
\usepackage{doi}
\usepackage{amsmath}
\usepackage{algorithm}
\usepackage{algorithmicx}
\usepackage[noend]{algpseudocode}
\usepackage{caption,subcaption}
\allowdisplaybreaks
\algnewcommand{\LeftComment}[1]{\Statex \(\triangleright\) #1}
\usepackage{stmaryrd}
\usepackage{color,soul}
\usepackage{wrapfig}
\usepackage{cite}
\usepackage{cancel}
\hypersetup{
    colorlinks=true,
    linkcolor=blue,
    filecolor=magenta,      
    urlcolor=cyan,
}
\usepackage{courier}

\usepackage{listings}
\usepackage{xcolor}
\definecolor{codegreen}{rgb}{0,0.6,0}
\definecolor{codegray}{rgb}{0.5,0.5,0.5}
\definecolor{codepurple}{rgb}{0.58,0,0.82}
\definecolor{backcolour}{rgb}{0.95,0.95,0.92}

\lstdefinestyle{mystyle}{
  backgroundcolor=\color{backcolour},   commentstyle=\color{codegreen},
  keywordstyle=\color{magenta},
  numberstyle=\tiny\color{codegray},
  stringstyle=\color{codepurple},
  basicstyle=\ttfamily\footnotesize,
  breakatwhitespace=false,         
  breaklines=true,                 
  captionpos=b,                    
  keepspaces=true,                 
  numbers=left,                    
  numbersep=5pt,                  
  showspaces=false,                
  showstringspaces=false,
  showtabs=false,                  
  tabsize=2
}

\title{An efficient algorithm for numerical homogenization of fluid filled porous solids: part-I}

\author{
  {
  Saumik Dana}\\
	University of Southern California\\
	Los Angeles, CA 90007 \\
	\texttt{sdana@usc.edu} \\
 		\And
  {
  Mary F Wheeler} \\
	Oden Institute for Computational Engineering and Sciences\\
	University of Texas at Austin\\
	Austin, TX 78712 
}

\date{}

\begin{document}
\maketitle
\begin{abstract}
The concept of representative volume element or RVE is invoked to develop an algorithm for numerical homogenization of fluid filled porous solids. RVE based methods decouple analysis of a composite material into analyses at the local and global levels. The local level analysis models the microstructural details to determine effective properties by applying boundary conditions to the RVE and solving the resultant boundary value problem. The composite structure is then replaced by an equivalent homogeneous material having the calculated effective properties. 
We combine the features of two techniques: one is the definition of a displacement field for the fluid phase to allow for a definition of a continuous displacement field across the microstructure and the other is the $FE^2$ numerical homogenization that couples the macroscale with the RVE scale via gauss points.
\end{abstract}
\section{Introduction}
\begin{figure}[htb!]
\centering
\includegraphics[scale=0.85]{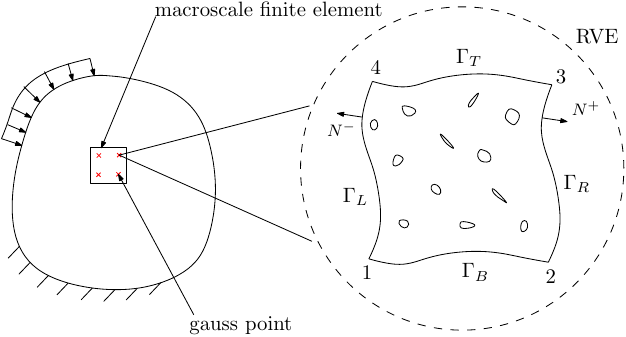}
\caption{A 2D depiction of the algorithmic framework. The macroscale boundary value problem is discretized into finite elements. The gauss point level computations for the macroscale BVP work in conjunction with RVE scale solve corresponding to each gauss point.}
\label{dep}
\end{figure}
The RVE concept~\cite{hashin-1962,hill-1963,hillmandel1,hill-1972,hashin-1983,zohdi,terada-2001,fishbook} is commonly used in the aerospace/automotive industry to avoid using computationally expensive simulation platforms necessary to capture microstructural features. In essence, the features are captured in the RVE and averaged out over the RVE before any discretization technique is employed at the macroscale with the averaged properties as parameters. More often than not, a number of simulations are run with different microstructures and the statistical mean of the results from those simulations on the macroscale are used as guiding principles for the design of the aerospace/automotive part. The reason for running multiple simulations each with a different microstructure is that the microstructure is only known stochastically and not deterministically. With advances in the field of material science and increased emphasis on modeling of biological tissues and engineered structures like foams and textiles, the need is felt to extend the analyses for cases in which both solid and fluid phases exist in tandem in the microstructure. The problem is complicated since the physics of solid dictates focus on solid displacements while the physics of fluid dictates focus on fluid velocities. The displacements and velocities are dimensionally incompatible since the velocity represents the rate of change of fluid displacement. In lieu of that, we employ the technique in which a displacement field is defined for the fluid phase also to go with the displacement field defined naturally for the solid phase~\cite{sandstrom-2016}. We combine that technique with the $FE^2$ homogenization framework~\cite{geers,ozdemir-2008,schroder-2014} in which each gauss point for the finite element calculations at the macroscale is associated with a RVE and the information exchange between the two scales occurs at each of those gauss points via the deformation gradient. A 2D depiction of the algorithmic framework is given in Figure \ref{dep}. The reason for calling the framework $FE^2$ is that both the macroscale and the RVE scale are solved using finite element method. 
The accuracy of the RVE approximation depends on how well the assumed boundary conditions reflect each of the myriad boundary conditions to which the RVE is subjected in-situ. The imposed boundary conditions on the RVE should be such that the Hill-Mandel condition~\cite{hillmandel1,hazanov-1998,hazanov-1999,mune-1999,fish-2008,temizer-2011,peric-2011,larsson-2011a,saeb-2016} of energetic equivalence between the two scales is satisfied. Periodic boundary conditions satisfy the Hill-Mandel condition and are generally the optimal choice from the standpoint of macroscale accuracy~\cite{swan-1994,michel-1999,takano-2000,kouznetsova2001,miehe-2002,miehe-2002a,miehe-2003,temizer-2007,temizer-2008a,temizer-2011a,larsson-2011,sengupta-2012,dijk-2015,vaz-2010}. It is important to note that this framework is meant for monolithic solution of the coupled system of equations as opposed to the popular fixed stress split strategy~\cite{dana-2018,dana2019design,dana2019system,dana2020,dana2021,danacg,danacmame,danathesis} in which the coupled set of equations are first decoupled and then solved sequentially and iteratively until convergence at each time step. This paper is structured as follows: we explain the concepts of deformation gradient and first P-K stress in this Section. We present the algorithmic framework for numerical homogenization in purely mechanical case in Section 2. We then proceed to present the algorithmic framework in the hydromechanical case of fluid filled porous solids in Section 3. We finally present conclusions and outlook in Section 4. An important part of the derivation of the RVE level system of equations for the fluid filled porous solid case with be dealt with in part-II of this paper.
\subsection{The deformation gradient and first P-K stress}\label{explain}
\begin{figure}[htb!]
\centering
\includegraphics[scale=0.9]{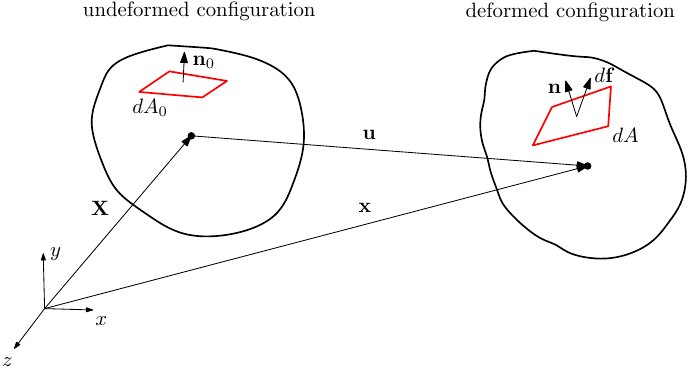}
\caption{$\mathbf{X}$ is position vector of point in reference configuration and $\mathbf{x}=\mathbf{X}+\mathbf{u}$ is the position vector the same point in the deformed configuration. Meanwhile, an elemental area $dA_0$ with unit normal $\mathbf{n}_0$ deforms to $dA$ with unit normal $\mathbf{n}$ under the transformation.}
\label{defgrad}
\end{figure}
As shown in Figure \ref{defgrad}, let $\mathbf{u}$ be the macroscale deformation field. 
The macroscale deformation gradient $\mathbf{F}_M$ is the macroscale spatial derivative of $\mathbf{x}$ in the reference configuration as follows
\begin{align*}
\mathbf{F}=\mathbf{x} \otimes \nabla_X \equiv \mathbf{I}+\mathbf{u}\otimes \nabla_X
\end{align*}
 An incremental force $d\mathbf{f}$ is defined with respect to the Cauchy stress $\boldsymbol{\sigma}$ and the first Piola-Kirchoff stress $\mathbf{P}$ in the deformed and reference configurations respectively as follows
\begin{align*}
d\mathbf{f}=\boldsymbol{\sigma}\mathbf{n}\,dA=\mathbf{P}\mathbf{n}_0\,dA_0
\end{align*}
\section{Algorithmic framework for the pure mechanical case}
In the initialization stage,
\begin{itemize}
\item The macroscale deformation gradient is set to identity tensor since that would imply the macroscale BVP starts at the reference configuration
\item The macroscale BVP is discretized into finite elements and an RVE is assigned to each gauss point
\item The macroscale tangent stiffness is not known apriori and is obtained at each gauss point from RVE level computations as shown in module \ref{macro}
\end{itemize}
Once the initalization phase is complete, 
\begin{enumerate}
\item An increment of macro load is applied
\item Macroscale BVP is solved
\item The macroscale deformation gradient is updated
\item Periodic boundary conditions are imposed on RVE in accordance with \eqref{perbc}
\item RVE BVP is solved and homogenized first P-K stress is obtained in accordance with \eqref{four}
\item The gauss point level homogenized first P-K stress is used to compute internal forces at macroscale finite element nodes
\end{enumerate}
If these internal forces are in balance with the prescribed macro load, incremental convergence has been achieved and steps $1-6$ are repeated. If that is not the case, steps $2-6$ are repeated. 
\subsection{Periodic boundary conditions on RVE}\label{perbcmod}
\begin{figure}[htb!]
\centering
\includegraphics[scale=1.0]{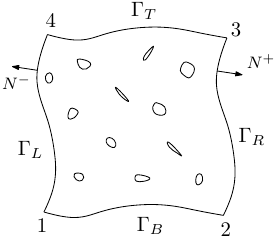}
\caption{Typical 2D RVE with pertinent microstructural features. $\Gamma_L$ and $\Gamma_R$ are mirror images so are $\Gamma_T$ and $\Gamma_B$. This helps in easy implementation of periodic boundary conditions on the RVE in accordance with \cite{kouznetsova2001}.}
\label{per}
\end{figure}
The typical RVE for imposition of periodic boundary conditions is shown in Figure \ref{per}. After each macroscale BVP solve, the deformation gradient is updated and the new position vectors of the vertices of the RVE are obtained using
\begin{align}
\label{perbc}
\mathbf{x}=\mathbf{F}_M \mathbf{X}
\end{align}
where $\mathbf{X}$ represents position vector in the reference configuration. This alongwith the shape periodicity of the RVE enables the implementation of periodic boundary conditions on RVE. It is easy to see that the prescribed periodic boundary conditions are Dirichlet boundary conditions. 
\subsection{Computation of homogenized tangent stiffness at the macroscale}\label{macro}
The linear momentum balance for the macroscale BVP in the reference configuration is given by
\begin{align*}
\nabla_X \cdot \mathbf{P}_M+\mathbf{b}=\mathbf{0}
\end{align*}
where $\mathbf{b}$ is the body force vector. The macroscale incremental constitutive law is
\begin{align}
\label{zero}
\delta \mathbf{P}_M=\mathbb{C}_M\delta \mathbf{F}_M
\end{align}
where $\mathbb{C}_M$ is the fourth order macroscale material property tensor.
The determination of $\mathbb{C}_M$ proceeds as follows: First, the RVE scale linear momemtum balance is expressed in the indicial notation as
\begin{align*}
P_{ik,k}+b_i=0\qquad i,k=1,2,3
\end{align*}
where the notation $(\cdot)_{,k}$ is used to denote the spatial derivative in the reference configuration as follows 
\begin{align*}
(\cdot)_{,k}\equiv \frac{\partial (\cdot)}{\partial X_k}
\end{align*}
Before we proceed, we assume that the body force is zero, and obtain the following using chain rule for differentiation 
\begin{align}
\label{one}
(P_{ik}X_j)_{,k}=P_{ik,k}X_j+P_{ik}\delta_{jk}=-\cancelto{0}{b_i} X_j+P_{ij}
\end{align}
We express the macroscale first P-K stress in indicial notation as follows
\begin{align}
\label{two}
P_{M_{ij}}=\frac{1}{V_0}\int\limits_{V_0}P_{ij}\,dV_0=\frac{1}{V_0}\int\limits_{V_0}(P_{ik}X_j)_{,k}\,dV_0=\frac{1}{V_0}\int\limits_{\Gamma_0}P_{ik}n_{0_k} X_j \,d\Gamma_0
\end{align}
where the third equality follows from \eqref{one} and the fourth equality follows from divergence theorem.
We then write \eqref{two} in tensorial notation as
\begin{align}
\label{three}
\mathbf{P}_M=\frac{1}{V_0}\int\limits_{\Gamma_0}\mathbf{t}_0\otimes \mathbf{X}\,d\Gamma_0
\end{align}
We know that the RVE level BVP is also solved using finite elements. Let $N_p$ be the number of boundary nodes for the RVE scale discretized domain and let $\mathbf{f}_p^{(i)}$ be the force on $i^{th}$ boundary node. We can rewrite \eqref{three} as 
\begin{align}
\label{four}
\mathbf{P}_M=\frac{1}{V_0}\int\limits_{\Gamma_0}\mathbf{t}_0\otimes \mathbf{X}\,d\Gamma_0=\frac{1}{V_0}\sum\limits_{i=1}^{N_p}\mathbf{f}_p^{(i)} \otimes \mathbf{X}^{(i)}
\end{align}
It is important to note that these boundary forces are not known apriori and are obtained from the RVE level solve.
Let $\mathbf{u}_f$ represent the displacement DOFs corresponding to the interior nodes and $\mathbf{u}_p$ represent the displacement DOFs corresponding to the boundary nodes. The force displacement relation for the RVE scale problem is
\begin{align*}
\overbrace{\begin{bmatrix}
\mathbf{K}_{pp} & \mathbf{K}_{pf}\\
\mathbf{K}_{fp} & \mathbf{K}_{ff}
\end{bmatrix}}^{\mathbf{K}^{RVE}}\left\{\begin{array}{c}
\delta \mathbf{u}_p\\
\delta \mathbf{u}_f
\end{array} \right\}=\left\{\begin{array}{c}
\delta \mathbf{f}_p\\
\mathbf{0}
\end{array} \right\}
\end{align*}
where the matrix $\mathbf{K}^{RVE}$ is dictated by the microstructure and is known apriori. We knock off DOFs corresponding to internal nodes to obtain
\begin{align}
\label{five}
[\overbrace{\mathbf{K}_{pp}-\mathbf{K}_{pf}(\mathbf{K}_{ff})^{-1}\mathbf{K}_{fp}}^{\mathbf{K}}]\{\delta \mathbf{u}_p\}=\{\delta \mathbf{f}_p\}
\end{align}
The incremental macroscopic first PK stress is obtained as
\begin{align}
\nonumber
\delta \mathbf{P}_M&=\frac{1}{V_0}\sum\limits_{i=1}^{N_p}\delta \mathbf{f}_p^{(i)} \otimes \mathbf{X}^{(i)}\qquad (from\,\eqref{four})\\
\nonumber
&=\frac{1}{V_0}\sum\limits_{i=1}^{N_p}\sum\limits_{j=1}^{N_p}\mathbf{K}^{(ij)}\delta \mathbf{u}_p^{(j)} \otimes \mathbf{X}^{(i)}\qquad (from\,\eqref{five})\\
\label{six}
&=\frac{1}{V_0}\sum\limits_{i=1}^{N_p}\sum\limits_{j=1}^{N_p}\mathbf{K}^{(ij)}\delta \mathbf{F}_M \mathbf{X}^{(j)}\otimes \mathbf{X}^{(i)}\qquad (\delta \mathbf{u}=\delta \mathbf{F}_M \mathbf{X})
\end{align}
Comparing \eqref{six} with \eqref{zero}, we get
\begin{align}
\label{seven}
\mathbb{C}_{M_{abcd}}= \frac{1}{V_0}\sum\limits_{i=1}^{N_p}\sum\limits_{j=1}^{N_p}\mathbf{K}_{ac}^{(ij)}\mathbf{X}^{(i)}_b\mathbf{X}^{(i)}_d\qquad a,b,c,d=1,2,3
\end{align}
\begin{algorithm}[H]
\begin{algorithmic}
\caption{Pure mechanical case}
\State $\mathbf{F}_M\gets \mathbf{I}$ \Comment Initialize deformation gradient
\For{$E \in \mathscr{T}_h$} \Comment loop over macroscale finite elements
\For{$g \in \mathscr{G}$}\Comment loop over gauss points
\State RVE $\leftrightarrow\,\,g$ \Comment Assign a RVE to each gauss point
\State Discretize the RVE
\State Calculate homogenized macroscopic tangent stiffness in accordance with \eqref{seven} and store it
\EndFor
\State Assemble macroscopic tangent stiffness over gauss points
\EndFor
\State Assemble macroscopic tangent stiffness over finite elements
\While{$t\le T$}
\State Apply increment of macro load
\While{(Internal force-Macro load $>$ TOL)}
\State Solve macroscale problem for $\delta \mathbf{F}_M$
\State $\mathbf{F}_M\gets \mathbf{F}_M+\delta \mathbf{F}_M$ \Comment Update deformation gradient
\For{$E \in \mathscr{T}_h$} \Comment loop over macroscale finite elements
\For{$g \in \mathscr{G}$}\Comment loop over gauss points
\State Prescribe periodic BCs in accordance with \eqref{perbc} 
\State Solve RVE problem
\State Calculate first P-K stress in accordance with \eqref{four}
\EndFor
\EndFor
\State Compute internal forces at finite element nodes
\EndWhile
\EndWhile
\end{algorithmic}
\end{algorithm}
\section{Algorithmic framework for fluid filled porous solid case}
\begin{figure}[htb!]
\centering
\includegraphics[scale=1]{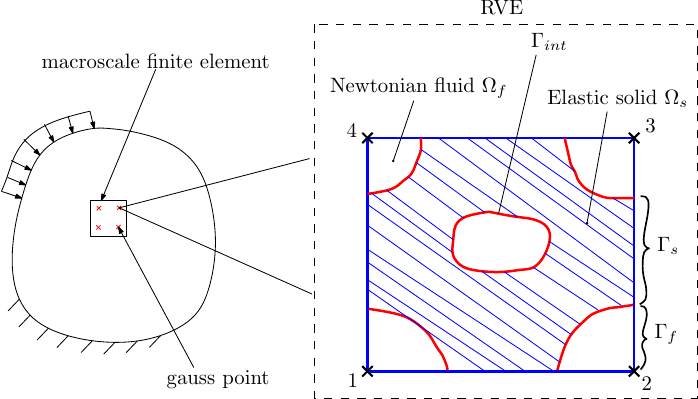}
\caption{A 2D depiction of the algorithmic framework. The macroscale boundary value problem is discretized into finite elements. The gauss point level computations for the macroscale BVP work in conjunction with RVE scale solve corresponding to each gauss point. Many pore morphology posbilities exist and we consider only one of those possibilities here.}
\label{solflu}
\end{figure}
In accordance with \cite{sandstrom-2016} and as shown in Figure \ref{solflu}, the solid phase inside the RVE is located in $\Omega_s$ and the fluid phase is located in $\Omega_f$. The interface between the two phases is $\Gamma_{int}$. A fictitious elastic material is introduced in $\Omega_f$. The material is then attached to the internal boundary $\Gamma_{int}$. As the interface moves, the fictitious elastic material is deformed, and the nodes inside $\Omega_f$ follow the deformation of the solid. Let $\Gamma=\Gamma_f\cup \Gamma_s$ where $\Gamma_f$ is the part of $\Omega_f$ where fluid can enter or exit the domain and $\Gamma_s$ is the part of $\Omega_s$ on the outer boundary.  
\subsection{Variables involved}
\begin{itemize}
\item RVE scale displacement field $\mathbf{u}=
\left\{\begin{array}{c}
\mathbf{u}_s\,\,\mathrm{in}\,\,\Omega_s, \\
\mathbf{u}_f\,\,\mathrm{in}\,\,\Omega_f,
\end{array}\right\}$
\item RVE scale first P-K stress $\mathbf{P}=
\left\{\begin{array}{c}
\mathbf{P}^s\,\,\mathrm{in}\,\,\Omega_s, \\
\mathbf{P}^f\,\,\mathrm{in}\,\,\Omega_f,
\end{array}\right\}$
\item Macroscale displacement field $\tilde{\mathbf{u}}$
\item Macroscale pressure field $\tilde{p}$
\item RVE scale fluid pressure field $p$ in $\Omega_f$
\item RVE scale fluid velocity field $\mathbf{v}$ in $\Omega_f$
\item RVE scale relative fluid-solid velocity field $\mathbf{w}\equiv \mathbf{v}-\frac{d\mathbf{u}_f}{dt}$ on $\Gamma_{int}$
\end{itemize}
\subsection{RVE level equations}
In accordance with \cite{sandstrom-2016}, the analysis is restricted to that of laminar and incompressible flow without convective acceleration i.e. Stokes' flow. Let $\mathbf{F}=\mathbf{I}+\mathbf{u}_f\otimes \nabla_X$ be the deformation gradient of the fictitious elastic material and let $J=det\,\mathbf{F}$. The fluid-structure interaction problem in the reference configuration is
\begin{equation*}
\left.\begin{array}{c}
\nabla \cdot \mathbf{P}^s=0\,\,in\,\,\Omega_s\qquad(\mathrm{linear\,\,momentum\,\,balance\,\,of\,\,solid\,\,phase})\\
\nabla \cdot \mathbf{P}^f=0\,\,in\,\,\Omega_f\qquad(\mathrm{linear\,\,momentum\,\,balance\,\,of\,\,fictitious\,\,solid})\\
J\mathbf{F}^{-T}:[\mathbf{v}\otimes \nabla]=0\,\,in\,\,\Omega_f\qquad(\mathrm{mass\,\,conservation\,\,of\,\,fluid\,\,phase})
\end{array}\right\}
\end{equation*}
On the internal boundary $\Gamma_{int}$, we have
\begin{itemize}
\item $\mathbf{P}_s\mathbf{n}+\mathbf{P}_f\mathbf{n}=\mathbf{0}$
\item $\mathbf{u}_f=\mathbf{u}_s$
\item $\mathbf{w}=\mathbf{0}$
\end{itemize}
\subsection{Periodic boundary conditions on RVE}
Let $\mathbf{X}_c$ be position vector of center point of RVE in reference configuration, $\tilde{p}\vert_c$ be macroscale pressure field evaluated at $\mathbf{X}_c$ and $(\nabla_X \tilde{p})\vert_c$ be the macroscale spatial gradient of $\tilde{p}$ evaluated at $\mathbf{X}_c$. The periodic boundary conditions are
\begin{itemize}
\item $\mathbf{x}=\mathbf{F}_M\mathbf{X}=(\mathbf{I}+\nabla_X \tilde{\mathbf{u}})\,\mathbf{X}\qquad \forall\,\,RVE\,\,vertices$
\item $p=\tilde{p}\vert_c+(\nabla_X \tilde{p})\vert_c\cdot (\mathbf{X}-\mathbf{X}_c),\qquad \forall\,\,RVE\,\,vertices$
\end{itemize}
\subsection{Computation of homogenized macroscopic tangent stiffness and first P-K stress}
The RVE level system of equations is obtained as
\begin{equation}
\label{RVEeq}
\begin{bmatrix}
\mathbf{K}_{uu} & \mathbf{K}_{up}\\
\mathbf{K}_{pu} & \mathbf{K}_{pp}
\end{bmatrix}\left\{\begin{array}{c}
\delta \mathbf{u}\\
\delta p
\end{array}\right\}=\left\{\begin{array}{c}
\delta \mathbf{f}_s\\
\delta \mathbf{f}_p
\end{array}\right\}
\end{equation}
where $\mathbf{f}_s$ and $\mathbf{f}_p$ are resulting boundary forces corresponding to the solid and fluid phases respectively. The details of the derivation of \eqref{RVEeq} will be explained in part-II of this paper. The incremental macroscopic first PK stresses are obtained as
\begin{align*}
\left\{\begin{array}{c}
\delta \mathbf{P}_{s_M}\\
\delta \mathbf{P}_{f_M}
\end{array}\right\}=\left\{\begin{array}{c}
\frac{1}{\vert \Omega_s\vert}\int\limits_{\Omega_s}\delta \mathbf{P}_{s}\\
\frac{1}{\vert \Omega_f\vert}\int\limits_{\Omega_f}\delta \mathbf{P}_{f}
\end{array}\right\}=\left\{\begin{array}{c}
\frac{1}{\vert \Omega_s\vert}\int\limits_{\Gamma}\delta \mathbf{t}_{s}\otimes \mathbf{X}\\
\frac{1}{\vert \Omega_f\vert}\int\limits_{\Gamma}\delta\mathbf{t}_{f}\otimes \mathbf{X}
\end{array}\right\}=\left\{\begin{array}{c}
\frac{1}{\vert \Omega_s\vert}\sum\limits_{i=1}^4\delta \mathbf{f}_{s}\otimes \mathbf{X}^{(i)}\\
\frac{1}{\vert \Omega_f\vert}\sum\limits_{i=1}^4\delta \mathbf{f}_{p}\otimes \mathbf{X}^{(i)}
\end{array}\right\}
\end{align*}
which, in lieu of \eqref{RVEeq} can be written as
\begin{align*}
\left\{\begin{array}{c}
\delta \mathbf{P}_{s_M}\\
\delta \mathbf{P}_{f_M}
\end{array}\right\}=\left\{\begin{array}{c}
\frac{1}{\vert \Omega_s\vert}\sum\limits_{i=1}^4\sum\limits_{j=1}^4(\mathbf{K}_{uu_{ij}}\delta \mathbf{u}^{(j)}+\mathbf{K}_{up_{ij}}\delta p^{(j)})\otimes \mathbf{X}^{(i)}\\
\frac{1}{\vert \Omega_f\vert}\sum\limits_{i=1}^4\sum\limits_{j=1}^4(\mathbf{K}_{pu_{ij}}\delta \mathbf{u}^{(j)}+\mathbf{K}_{pp_{ij}}\delta p^{(j)})\otimes \mathbf{X}^{(i)}
\end{array}\right\}
\end{align*}
We then substitute $\delta \mathbf{u}^{(j)}=\delta \mathbf{F}_M \mathbf{X}^{(j)}$ and $\delta p^{(j)}=\delta \tilde{p}\vert_c+\delta (\nabla_X \tilde{p})\vert_c\cdot (\mathbf{X}^{(j)}-\mathbf{X}_c)$ in the above and compare with \eqref{compare} 
\begin{align}
\label{compare}
\left\{\begin{array}{c}
\delta \mathbf{P}_{s_M}\\
\delta \mathbf{P}_{f_M}
\end{array}\right\}=\mathbf{K}_M \left\{\begin{array}{c}
\delta \mathbf{F}_{M}\\
\delta \tilde{p}
\end{array}\right\}
\end{align}
to obtain macroscopic tangent stiffness $\mathbf{K}_M$
\begin{algorithm}[H]
\begin{algorithmic}
\caption{Fluid filled porous solid case}
\State $\mathbf{F}_M\gets \mathbf{I}$ \Comment Initialize deformation gradient
\For{$E \in \mathscr{T}_h$} \Comment loop over macroscale finite elements
\For{$g \in \mathscr{G}$}\Comment loop over gauss points
\State RVE $\leftrightarrow\,\,g$ \Comment Assign a RVE to each gauss point
\State Discretize the RVE
\State Calculate homogenized macroscopic tangent stiffness and store it
\EndFor
\State Assemble macroscopic tangent stiffness over gauss points
\EndFor
\State Assemble macroscopic tangent stiffness over finite elements
\While{$t\le T$}
\State Apply increment of macro load
\While{(Internal force-Macro load $>$ TOL)}
\State Solve macroscale problem for $\delta \mathbf{F}_M$
\State $\mathbf{F}_M\gets \mathbf{F}_M+\delta \mathbf{F}_M$ \Comment Update deformation gradient
\For{$E \in \mathscr{T}_h$} \Comment loop over macroscale finite elements
\For{$g \in \mathscr{G}$}\Comment loop over gauss points
\State Prescribe periodic BCs
\State Solve RVE problem
\State Calculate first P-K stress
\EndFor
\EndFor
\State Compute internal forces at finite element nodes
\EndWhile
\EndWhile
\end{algorithmic}
\end{algorithm}
\section{Conclusions and outlook}
We combined the features of two techniques in literature to design an algorithm for numerical homogenization of fluid filled porous solids. For the clarity of explanation, we first presented the algorithm for the pure mechanical case in which the entire microstructure is a solid phase. We then proceeded to explain the algorithm for the case in which the microstructure has both solid and fluid phases. The details of the derivation of the RVE level system of equations will be dealt with in part-II of this paper. 

%

\appendix
\bibliographystyle{unsrt}    
\bibliography{sample}

\end{document}